\documentclass[a4paper, 12pt]{article}
\usepackage[T2A]{fontenc} % кодировка
\usepackage[utf8]{inputenc} % кодировка исходного кода
\usepackage[english]{babel} % локализация и переносы
\usepackage[left=2cm,right=2cm,top=2cm,bottom=3cm,bindingoffset=0cm]{geometry}
\usepackage{indentfirst}
\usepackage{graphicx}
\graphicspath{}
\DeclareGraphicsExtensions{.pdf,.png,.jpg,.jpeg}

\usepackage[table,xcdraw]{xcolor}
\usepackage{booktabs}

\usepackage{hyperref}
\usepackage{physics}
\usepackage{amssymb, amsmath}

\usepackage{abstract}
\usepackage{pgfplots}
\usepackage{physics}
\usepackage{tikz}
\pgfplotsset{compat=1.15}
\usepackage{mathrsfs}
\usetikzlibrary{arrows}

\usepackage{mathtools}
\DeclarePairedDelimiter\autobracket{(}{)}
\newcommand{\br}[1]{\autobracket*{#1}}

\usepackage{empheq}
\newenvironment{eqw}{\begin{equation} \begin{aligned}}{\end{aligned}    \end{equation}}
\newenvironment{eqw*}{\begin{equation*} \begin{aligned}}{\end{aligned}    \end{equation*}}

\usepackage{subfig}
\usepackage{authblk}
\title{Energy levels of the second-harmonic Hamiltonian at large photon numbers}
\author{Boulat Nougmanov$^{1,2}$}
\date{%
    $^1$Russian Quantum Center, Skolkovo IC, Bolshoy Bulvar 30, bld. 1, Moscow, 121205, Russia\\%
    $^2$Moscow Institute of Physics and Technology, 9 Institutsky Lane, Dolgoprudny, 141701, Russia
}
\begin{document}
\maketitle
\begin{abstract}
We study the energy spectrum of the degenerate $\chi^{(2)}$ Hamiltonian describing second-harmonic generation and parametric down-conversion at large total excitation numbers. Owing to the conservation of the total excitation number, the spectral problem reduces to the diagonalization of finite-dimensional tridiagonal matrices. Following the approach of Alvarez and Alvarez-Estrada, we map this problem onto an effective one-dimensional Schr\"odinger equation with a double-well potential. We then derive an implicit quantization condition for the energy levels near the top of the potential barrier and obtain two explicit asymptotic formulas applicable in complementary spectral regions. The ranges of applicability of the resulting formulas are established, and their accuracy is compared with exact matrix diagonalization and with the conventional JWKB approximation. The proposed approach provides an accurate analytical description of the energy levels near the center of the spectrum, where the conventional approximation loses accuracy.
\end{abstract}

\section{Introduction}
Nonlinear optical processes of three-wave mixing play a key role in modern quantum optics. In particular, interactions of the $\chi^{(2)}$ type underlie phenomena such as second harmonic generation and parametric down-conversion, which are widely used to obtain nonclassical states of light \cite{mandel1996optical, walls2008quantum}. These processes provide a convenient platform for the study of fundamental quantum effects, and are also of practical importance in the problems of quantum information and quantum technologies \cite{kok2007linear, braunstein2005quantum}.

From a theoretical point of view, the degenerate $\chi^{(2)}$ interaction is described by a Hamiltonian that converts two quanta of one mode into one quantum of another mode and vice versa. The presence of an invariant makes it possible to significantly simplify the description of the system. In particular, the interaction Hamiltonian commutes with the total excitation number operator, which leads to the decomposition of the state space into independent subspaces with a fixed value of the corresponding quantum number. In each such subspace, the problem is reduced to finding the eigenvalues of the effective Hamiltonian acting in a finite-dimensional space. Thus, the initial task of spectral analysis is reduced to the study of a family of finite-dimensional matrices.

It is known that the Hamiltonian under consideration belongs to the class of quasi-exactly solvable models \cite{alvarez2002quasi}. In particular, there is a representation that allows it to be written as a tridiagonal matrix, which significantly simplifies both numerical computations and analytical treatment \cite{nikitin1991quantum}. In addition, the spectrum has symmetry with respect to zero, which reflects the internal properties of the interaction and the structure of the corresponding operators \cite{kreshchuk2019classical}.

Despite this, a complete analytical description of the spectrum remains a non-trivial task. Even with a fixed value of the total excitation number, the structure of energy levels is highly nontrivial, and obtaining explicit expressions for eigenvalues is difficult \cite{wu2003spectrum}. In particular, questions remain open related to the behavior of the spectrum at large total excitation number ($N\gg 1$) and the possibility of its compact description.

The purpose of this work is to derive asymptotic expressions for the energy levels of the Hamiltonian of the degenerate $\chi^{(2)}$ interaction. We focus on the structure of the spectrum and identify the energy levels that are most relevant to the dynamics of physically important initial states.

Previous semiclassical analysis of the second-harmonic-generation Hamiltonian was mainly based on the conventional JWKB approximation developed by Alvarez and Alvarez-Estrada. That approximation provides an accurate description of the levels associated with motion near the minima of the effective double-well potential, but loses accuracy for levels approaching the separatrix energy. In the present work, we apply the separatrix quantization condition derived by Bleher for a generic double-well potential to the effective Schrödinger equation corresponding to the degenerate $\chi^{(2)}$ Hamiltonian. After accounting for the parity selection rule and the specific indexing of the optical spectrum, we derive an implicit quantization condition and obtain two explicit asymptotic expressions for the energy levels, Eqs. \eqref{eq:ultrasmall_E} and \eqref{eq:normal_small_E}, valid in complementary low- and intermediate-\(\lvert E\rvert\) regimes.

The paper is organized as follows. Section \ref{sec:gen_knowledge} summarizes the general properties of the Hamiltonian. In Section \ref{sec:reduction_to_Shrodinger}, we reduce the spectral problem to an effective one-dimensional Schr\"odinger equation. In Section \ref{sec:parameter_area}, we identify the part of the spectrum most relevant to physically important initial states. In Section \ref{sec:small_energies}, we derive the near-separatrix quantization condition and the corresponding asymptotic formulas. Section \ref{sec:conclusion} summarizes the results.
\section{Basic properties of the Hamiltonian}\label{sec:gen_knowledge}
Consider the Hamiltonian of the degenerate $\chi^{(2)}$ interaction in the form:
\begin{eqw}
    \hat{H}_0 &= \hbar \omega \hat{N} + \hbar \gamma \hat{H}\\
    \hat{N} &= \hat{a}^{\dag} \hat{a} + 2 \hat{b}^{\dag} \hat{b}\\
    \hat{H} &= \hat{a}^2\hat{b}^{\dag} + \hat{a}^{\dag2}\hat{b},
\end{eqw}
where $\omega$ is the frequency of mode $a$. The operators $\hat{H}$ and $\hat{N}$ commute, which allows one to choose a common eigenbasis. If the eigenvalues of the operator $\hat{H}$ are found and equal to $E_n^N$, then the eigenvalues of the original Hamiltonian $H_0$ are given by: $\hbar\omega N+\hbar\gamma E_n^N$. Therefore, the problem reduces to finding the eigenvalues of the operator $\hat{H}$.

Energy levels are found as eigenvalues of a tridiagonal matrix of size $\br{[N/2]+1}\times\br{[N/2]+1}$ with a symmetric spectrum, where $[\dots]$ denotes the floor function. We choose the indexing such that $E_n^N$ increases monotonically with $n$ and satisfies
\begin{eqw}\label{eq:levels_symmetry}
    E_{-n}^N = -E_{n}^N.
\end{eqw}
This means that for even $[N/2]$, $n$ will run through all values from $-n_{\max}$ to $n_{\max}$ and there is $E_0^N= 0$ in the spectrum, and in the case of odd $[N/2]$, the values run through the same range except for $n=0$. The upper bound of $n_{\max}$ is generally written as:
\begin{eqw}
    n_{\max} = \lceil [N/2]/2 \rceil,
\end{eqw}
where $\lceil \dots \rceil$ denotes the ceiling function.
Having discussed this fact, we now derive only positive energy levels, since all other levels are found from the condition \eqref{eq:levels_symmetry}.
\section{Reduction to the one-dimensional Schr\"odinger equation}\label{sec:reduction_to_Shrodinger}
Alvarez and Alvarez-Estrada \cite{alvarez1995semiclassical} showed that each optical eigenvalue $E_n^N$ corresponds to a Schr\"odinger eigenvalue $-E_n^N$ in the equation
\begin{eqw}
    -\psi''(x) +\left(\frac{1}{16}x^6 - \br{\frac{N}{2}+\frac{3}{4}}x^2+E_{n}^N\right)\psi(x) = 0
\end{eqw}
The optical spectrum corresponds to the lowest $[N/2]+1$ states in the parity sector $\psi(-x)=(-1)^N\psi(x)$, which form the finite quasi-exactly solvable sector. The remaining Schr\"odinger eigenstates do not belong to the optical spectral problem.

We now depart from the scaling and notation used by Alvarez and introduce the following dimensionless parameters.
\begin{eqw}\label{eq:h_def}
    h =\frac{3/4}{N+3/2}
\end{eqw}
By scaling a coordinate $x\rightarrow2^{3/4}h^{-1/4}x$, we get the equation:
\begin{eqw}
    -h^2\psi''(x) + \br{4x^6-3x^2+\varepsilon_n}\psi(x) = 0,
\end{eqw}
where we have introduced:
\begin{eqw}\label{eq:varepsilon_def}
    \varepsilon_n = \frac{3\sqrt{6}E_n^N}{4\br{N+3/2}^{3/2}}
\end{eqw}
This corresponds to the Schr\"odinger equation with potential $U(x) = 4x^6 - 3x^2$, energy $-\varepsilon$, mass $1/2$ and the reduced Planck constant $h$.
\section{Relevant spectral range}\label{sec:parameter_area}
Optical nonlinearities are typically weak, so efficient frequency conversion often requires strong pumping. This motivates our focus on the large-excitation regime:
\begin{eqw}
    N \gg 1
\end{eqw}
Let us now consider two practically important processes: second harmonic generation and spontaneous parametric down-conversion. In the case of second harmonic generation, all the energy is concentrated in the $a$ mode, and in the case of SPDC, all the energy is concentrated in the $b$ mode. In this section, for the sake of clarity, we will consider Fock states rather than coherent ones as initial states.
\begin{figure}[!tbp]
    \centering
    \subfloat[]{\includegraphics[width=0.6\textwidth]{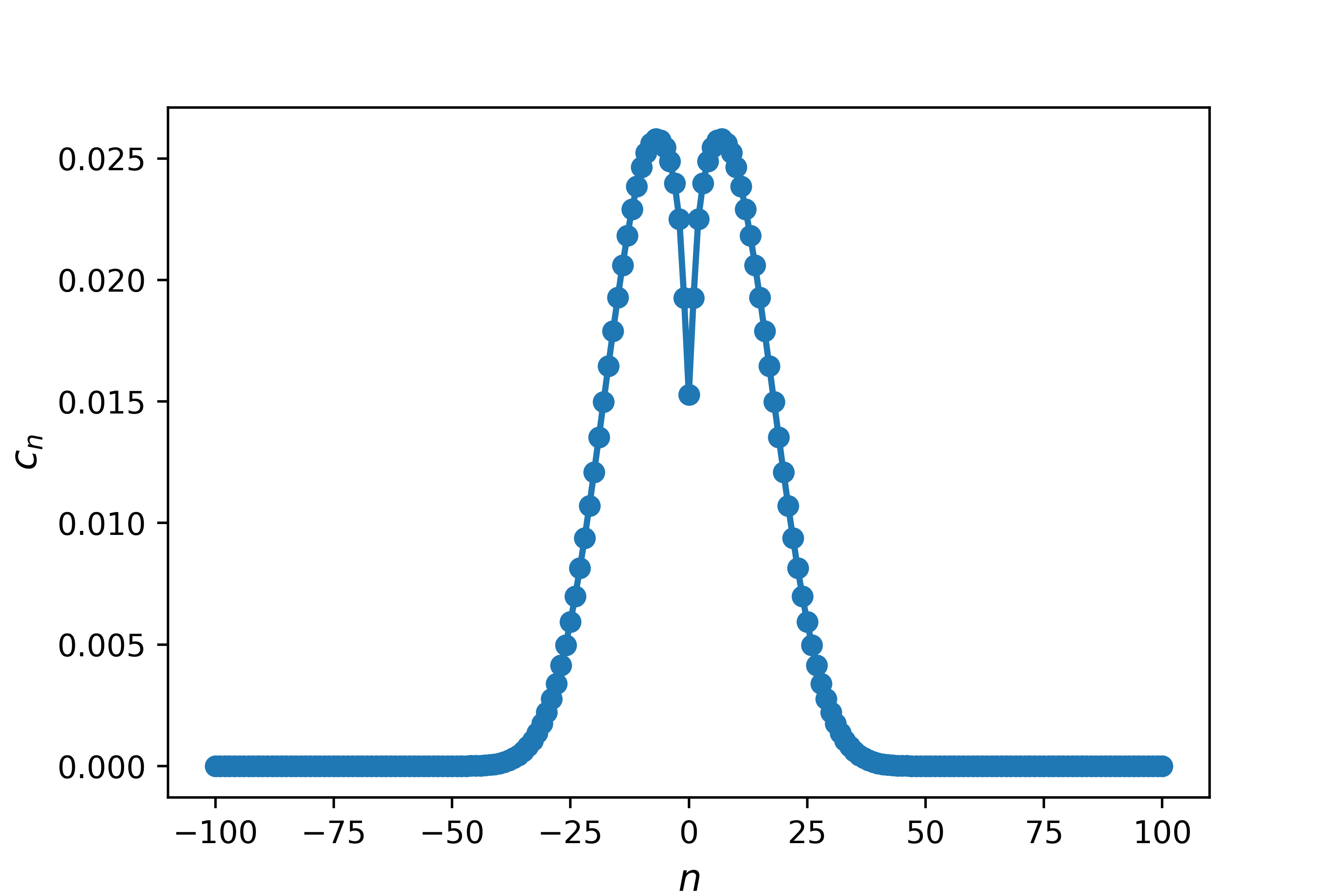}\label{fig:cna}}
    \hfill
    \subfloat[]{\includegraphics[width=0.6\textwidth]{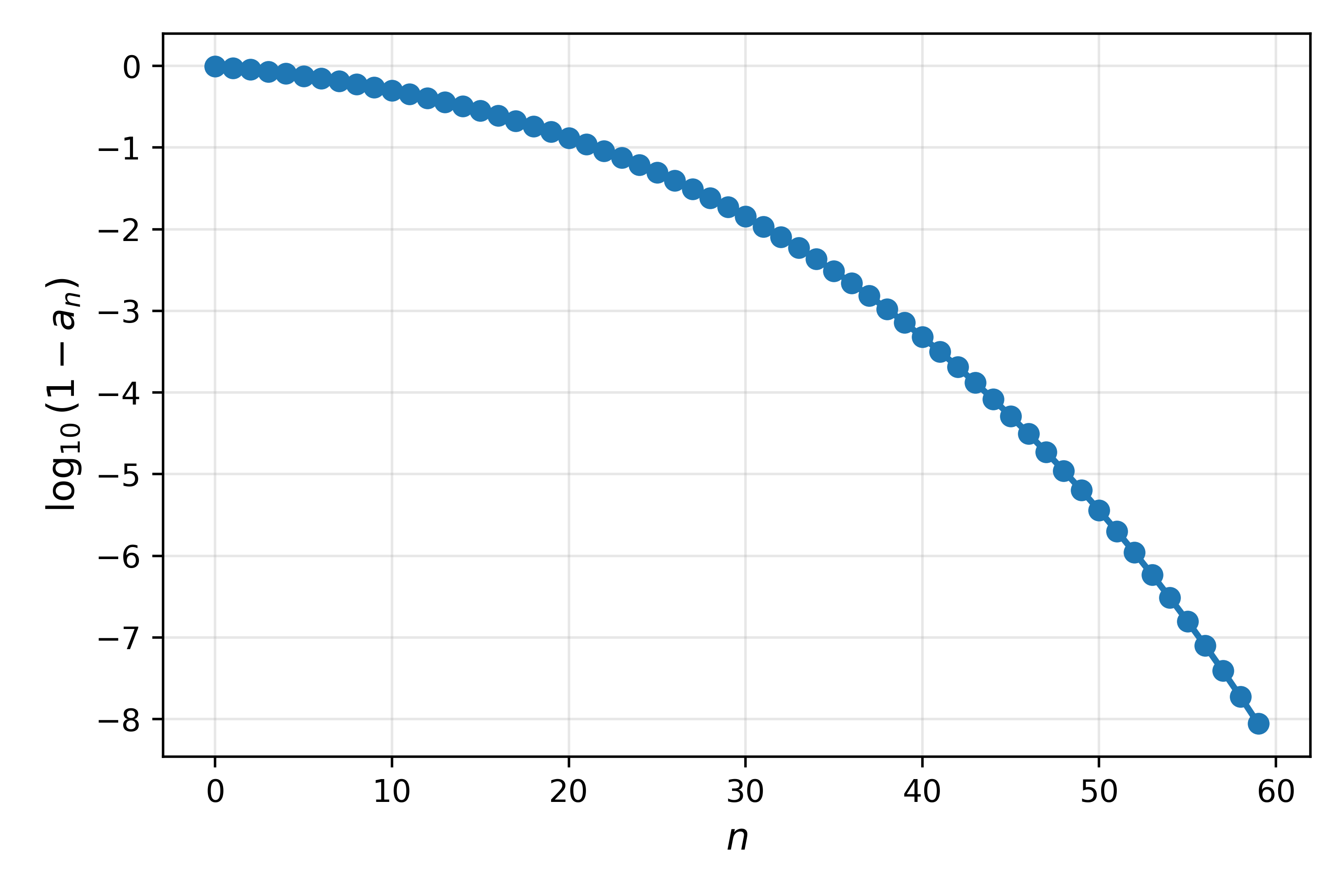}\label{fig:ana}}
    \caption{Expansion coefficients $c_n$ (a) and the accuracy of the truncated eigenstate expansion $a_n$ (b) for the initial state $\ket{\psi_0}=\ket{N}_a\ket{0}_b$.}
\end{figure}

To determine which energy levels make the largest contribution to the initial state, we introduce the expansion coefficients and the accuracy of the truncated eigenstate expansion:
\begin{eqw}
    c_k &= \abs{\braket{\psi_0}{\chi_k^N}},\\
    a_n &= \sum_{\abs{k}\leq\abs{n}} {c_k}^2
\end{eqw}
where $\ket{\chi_n^N}$ is the eigenstate of the Hamiltonian $\hat{H}$
corresponding to the energy level $E_n^N$. Figure \ref{fig:cna} shows the
expansion coefficients $c_n$ for the initial state associated with
second-harmonic generation, whereas Fig. \ref{fig:ana} shows the accuracy of
the expansion truncated to the central energy levels. The plotted distributions show that second-harmonic generation requires a broader set of central levels than spontaneous parametric down-conversion. The required cutoff depends on $N$ and on the desired retained weight $a_n$.
\begin{figure}[!tbp]
    \centering
    \subfloat[]{\includegraphics[width=0.6\textwidth]{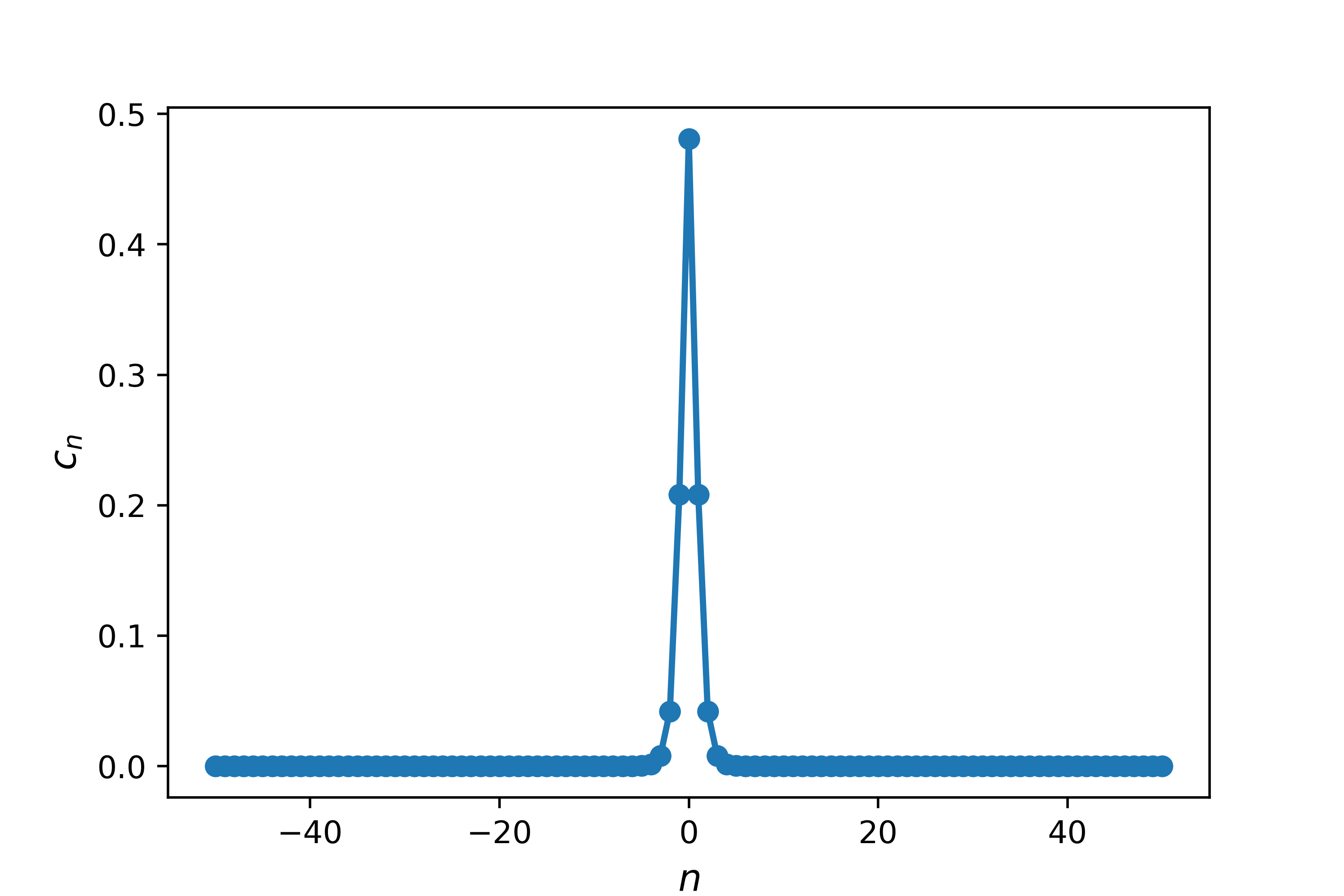}\label{fig:cnb}}
    \hfill
    \subfloat[]{\includegraphics[width=0.6\textwidth]{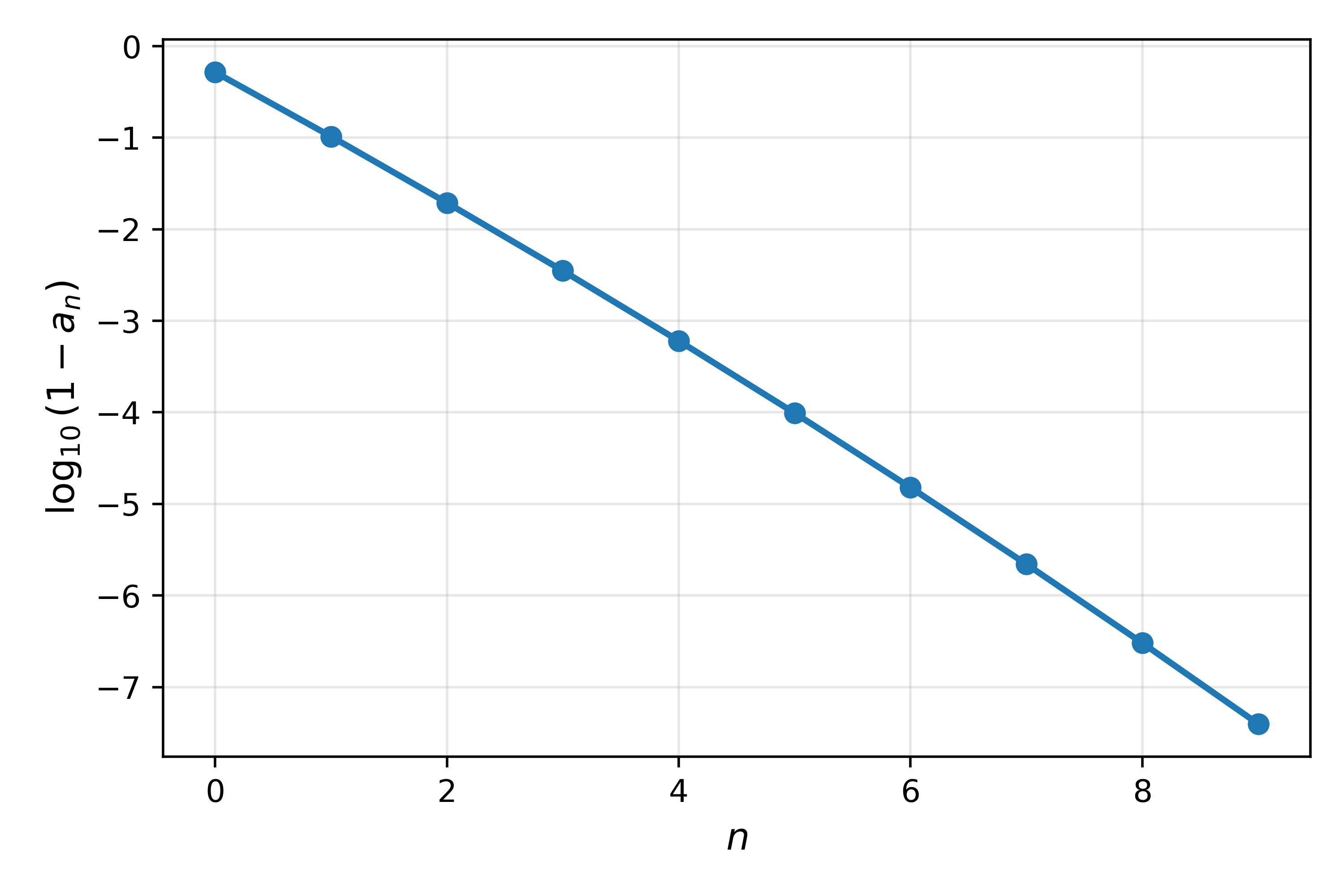}\label{fig:anb}}
    \caption{Expansion coefficients $c_n$ (a) and the accuracy of the truncated eigenstate expansion $a_n$ (b) for the initial state $\ket{\psi_0}=\ket{0}_a\ket{N/2}_b$.}
\end{figure}

Figures \ref{fig:cnb} and \ref{fig:anb} show the corresponding distributions for spontaneous parametric down-conversion, with the initial state $\ket{0}_a\ket{N/2}_b$ and even $N$. In this case, a small number of central levels carries most of the spectral weight for the values of $N$ shown. This concentration near the center of the spectrum was also observed in Ref.~\cite{gorshenin2025photon}. These numerical observations do not specify an asymptotic scaling of the cutoff with $N$.

Since the physically relevant initial states considered above have significant overlap mainly with the central, low-$\lvert E\rvert$ part of the spectrum, deriving accurate asymptotic expressions for these levels is of primary practical interest.

\section{Semiclassical quantization at small energies}\label{sec:small_energies}

\begin{figure}
    \centering
    \includegraphics[width=0.6\textwidth]{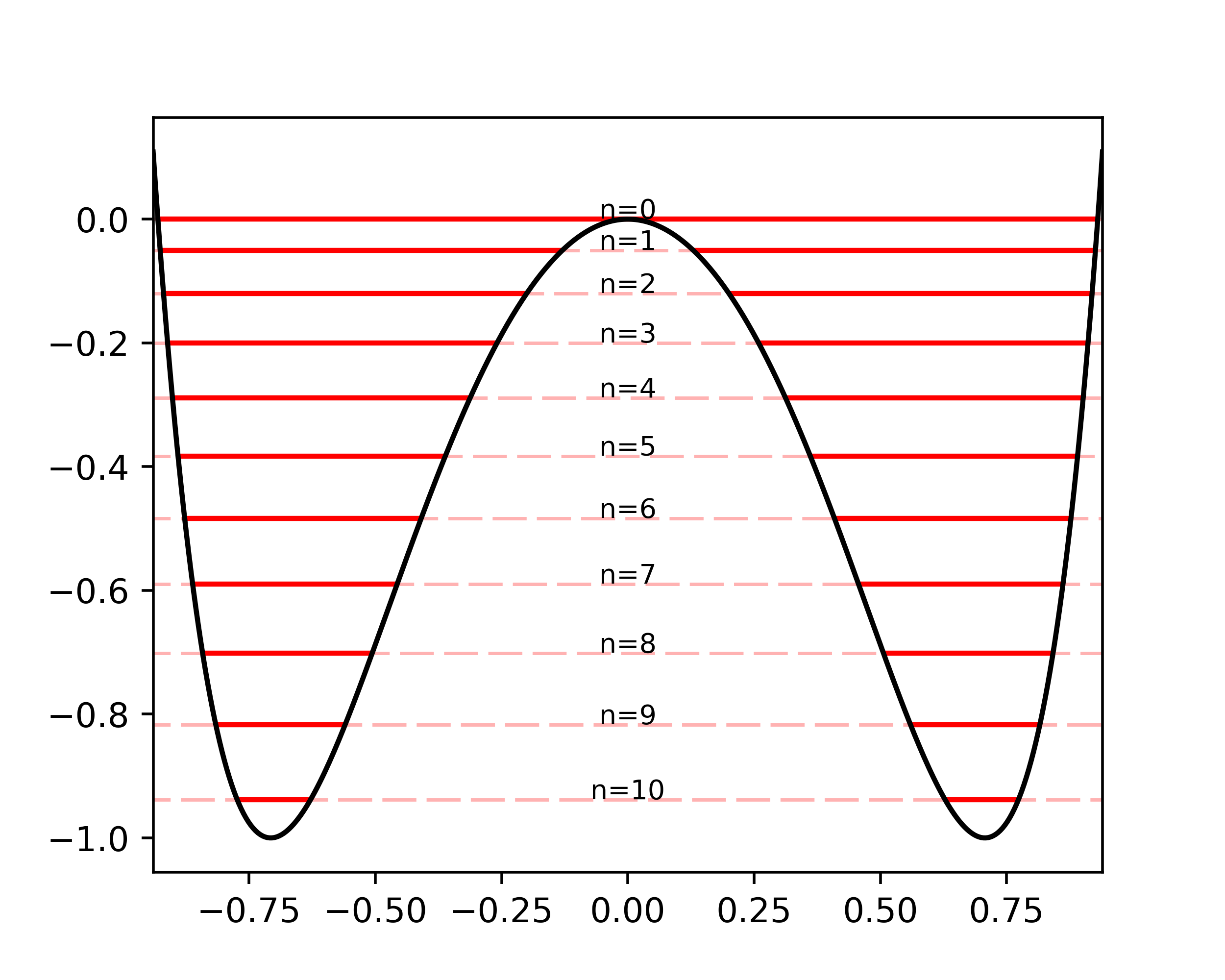}
    \caption{Potential values $U(x) = 4x^6 - 3x^2$ with indication of level lines $-\varepsilon_n$ at $N=40$.}
    \label{pic:potential}
\end{figure}
As shown in Fig. \ref{pic:potential}, these levels lie near the top of the barrier at $E=0$. In the corresponding classical system, this value is the separatrix energy. In this region, the conventional JWKB approximation loses accuracy, even when the tunneling correction introduced by Alvarez is taken into account \cite{alvarez1995semiclassical}. Bleher derived a quantization condition for this region that determines the scaled energies $\varepsilon_n$ with an error of order $O(h^2)$ \cite{bleher1994semiclassical}. To apply this condition to the present problem, we introduce the classical action integral:
\begin{eqw}\label{eq:I_def}
    I(\varepsilon) = \int\limits_{x_{\min}}^{x_{\max}}&\sqrt{-\varepsilon - 4x^6+3x^2} dx
\end{eqw}
We will also introduce another parameter, the range of which we will study later:
\begin{eqw}\label{eq:t_def}
    t = \frac{\varepsilon_n}{2\sqrt{3}h} = \frac{E_n^N}{\sqrt{2N+3}}
\end{eqw}
In the case of even $N$ we are only interested in even energy levels in the potential and vice versa. Then the equation for non-negative $\varepsilon_n$ is written as:
\begin{eqw}\label{eq:bleher_eq}
    \frac{I(\varepsilon_n)}{h} = \frac{1}{2}\br{\arg\br{\Gamma\br{\frac{1}{2} - i t}} +t \ln\abs{t} - t +(-1)^N \br{\frac{\pi}{2} + \arctan e^{\pi t}}} + \pi  (\tilde n - n)
\end{eqw}
\begin{eqw}
    \tilde{n} = \frac{N}{4} + \frac{1}{4}
    \left\{
    \begin{aligned}
        &0,\;\; N\underset{4}{\equiv} 0 \\
        &3,\;\; N\underset{4}{\equiv} 1\\
        &2,\;\; N\underset{4}{\equiv} 2\\
        &5,\;\; N\underset{4}{\equiv} 3
    \end{aligned}
    \right. 
    ,
\end{eqw}
In this formula, $\tilde n$ is introduced, depending on the remainder of $N$ when divided by 4. Such designations are related to the fact that in the original Bleher formula, the levels are numbered in pairs, starting from the lowest level. Appendix \ref{appendix_I_asymptotics} shows the asymptotics of the integral $I(\varepsilon)$:
\begin{eqw}\label{eq:I_asymptotics_till_5}
I(2\sqrt{3}\,ht)
&=
\frac{3\pi}{16}
+
\frac{ht}{2}
\left(
\ln(ht)-1-\ln 6
\right)+h \mathcal R(h,t)\\
\mathcal R(h,t)
&=
\frac{h^2t^3}{54}
\left(
15\ln(ht)+34-15\ln6
\right)
+
O\left(
h^4t^5\ln(ht)
\right)
\end{eqw}
Using the obtained asymptotic behavior, we can simplify expression \ref{eq:bleher_eq}. Appendix \ref{appendix_gamma_simplification} leads to the following expression:
\begin{eqw}\label{eq:bleher_without_zero_part}
    \frac t2\ln\frac3h
+
\arg\Gamma
\left(
\frac12-\frac{(-1)^N}{4}-\frac{it}{2}
\right)
=
\pi\kappa_n+\mathcal R(h,t)
\end{eqw}
where:
\begin{eqw}
    \kappa_n=
\begin{cases}
n, & N\equiv0,1\pmod4,\\[2mm]
n-\dfrac12, & N\equiv2,3\pmod4,
\end{cases}
\end{eqw}
Throughout, $\arg\Gamma$ denotes the continuous imaginary part of
$\log\Gamma$, normalized to zero at $t=0$. The shift $\kappa_n$
incorporates the level index and the parity-dependent phase correction.
For convenience, we define
\begin{eqw}\label{eq:central_parameters}
    D_N=\ln\frac{24}{h}+\gamma_{\mathrm E}+(-1)^N\frac\pi2,
\end{eqw}
where $\gamma_{\mathrm E}$ is Euler's constant. For $h\to0$ and
$\kappa_n=o(D_N)$, the small-$t$ expansion gives
\begin{eqw}\label{eq:ultrasmall_E}
    E_n^N \approx \sqrt{2N+3}\left(
    \frac{2\pi\kappa_n}{D_N}
    -\frac{\psi^{(2)}\!\left(\frac12-\frac{(-1)^N}{4}\right)(\pi\kappa_n)^3}{3D_N^4}
    \right).
\end{eqw}
Here $\psi^{(2)}$ is the second derivative of the digamma function.
Appendix~\ref{appendix_small_energies} derives this expression and its
remainder estimate.

For intermediate energies, we instead take $t\to\infty$ while
$ht\to0$. The leading Stirling approximation yields
\begin{eqw}\label{eq:normal_small_E}
    E_n^N \approx \sqrt{2N+3}\,
    \frac{2\pi\left(\kappa_n-\frac{(-1)^N}{8}\right)}
    {-W_{-1}\!\left(-\frac{\pi h}{3e}\left(\kappa_n-\frac{(-1)^N}{8}\right)\right)}.
\end{eqw}
Here $W_{-1}$ is the lower real branch of the Lambert $W$ function.
Appendix~\ref{appendix_medium_energies} gives the derivation, the
leading corrections, and the remainder estimates.
In the transition region $t=O(1)$, one should use the unexpanded
gamma phase in Eq.~\eqref{eq:bleher_without_zero_part}.
In conclusion, we will present the explicit formula used by Alvarez \cite{alvarez1995semiclassical} and valid for $\varepsilon\approx 1$:
\begin{eqw}\label{eq:alvarez_full}
    E_n^N \approx \br{\frac{2}{3}N+1}^{\frac{3}{2}}\br{1-2 \Delta\mathcal{E}-2\sum\limits_{l=0}^{\infty}\sum_{i=0}^{\infty}c_{2l}^{(i)}\br{\frac{2}{3}N+1}^{-2l-i}  \br{n_{\max}-n+\frac{1}{2}}^{i} }
\end{eqw}
\begin{eqw}
    \Delta\mathcal{E} = - \frac{2\sqrt{c(b-a)}}{\br{\frac{2}{3}N+1}\pi F\br{\frac{1}{2}, \frac{1}{2}; 1;\frac{a(b-c)}{c(b-a)}}}\exp\br{-2\pi\br{\frac{2}{3}N+1}b\sqrt{-ac}F_1\br{\frac{1}{2}, -\frac{1}{2}, -\frac{1}{2};2;\frac{b}{a},\frac{b}{c}}},
\end{eqw}
where $a,\;b, \;c$ are three roots of the polynomial $4 x^3 - 3 x + \br{\frac{2}{3}N+1}^{-3/2}E_n^N$ and $c_{2l}^{(i)}$ are special coefficients from the original paper \cite{alvarez1995semiclassical}.

The asymptotic regimes can be summarized in terms of the positive
level index as follows:
\begin{eqw}
    n\ll \ln N
    &\quad\Longrightarrow\quad
    \text{the small-$t$ approximation \eqref{eq:ultrasmall_E}},\\
    n\asymp\ln N
    &\quad\Longrightarrow\quad
    \text{the full gamma phase in \eqref{eq:bleher_without_zero_part}},\\
    \ln N\ll n\ll N
    &\quad\Longrightarrow\quad
    \text{the large-$t$ approximation \eqref{eq:normal_small_E}}.
\end{eqw}
The Alvarez approximation \eqref{eq:alvarez_full} applies near
$\varepsilon=1$. In the small-$t$ regime,
$t\sim2\pi\kappa_n/\ln N$, with the parity shift retained for fixed $n$.
In the intermediate regime \eqref{eq:normal_small_E}, the general leading behavior is instead
\begin{eqw}\label{eq:intermediate_scaling}
    t\sim\frac{2\pi n}{\ln(N/n)}.
\end{eqw}

Appendix~\ref{appendix_accuracy} gives the error estimates for the explicit approximations and the conditions for accuracy finer than the adjacent-level spacing.

\begin{figure}
    \centering
    \includegraphics[width=0.6\textwidth]{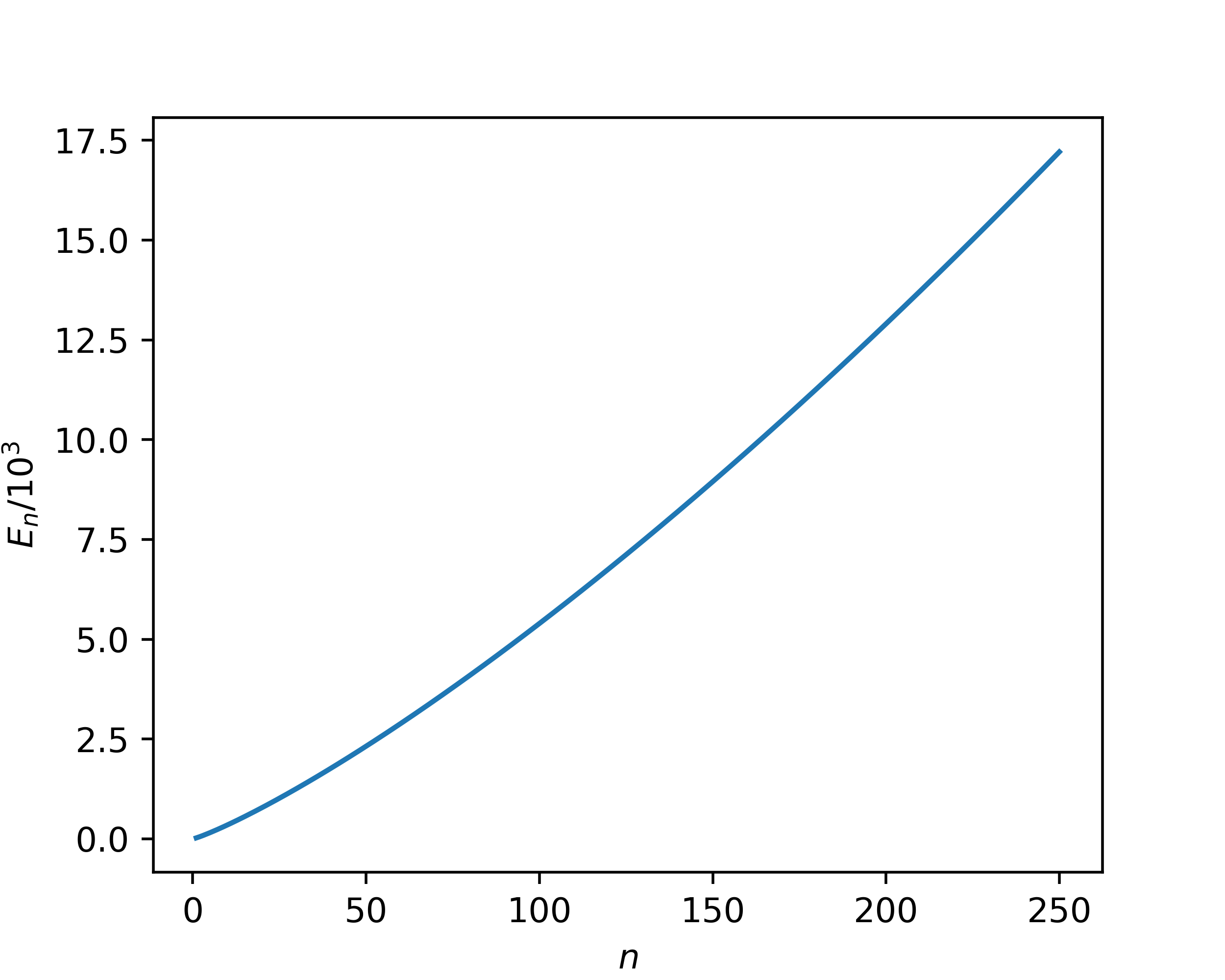}
    \caption{The exact energy levels found by the matrix diagonalization method in the Nikitin-Masalov representation at $N=1000$.}
    \label{pic:evals}
\end{figure}

\begin{figure}
    \centering
    \includegraphics[width=0.6\textwidth]{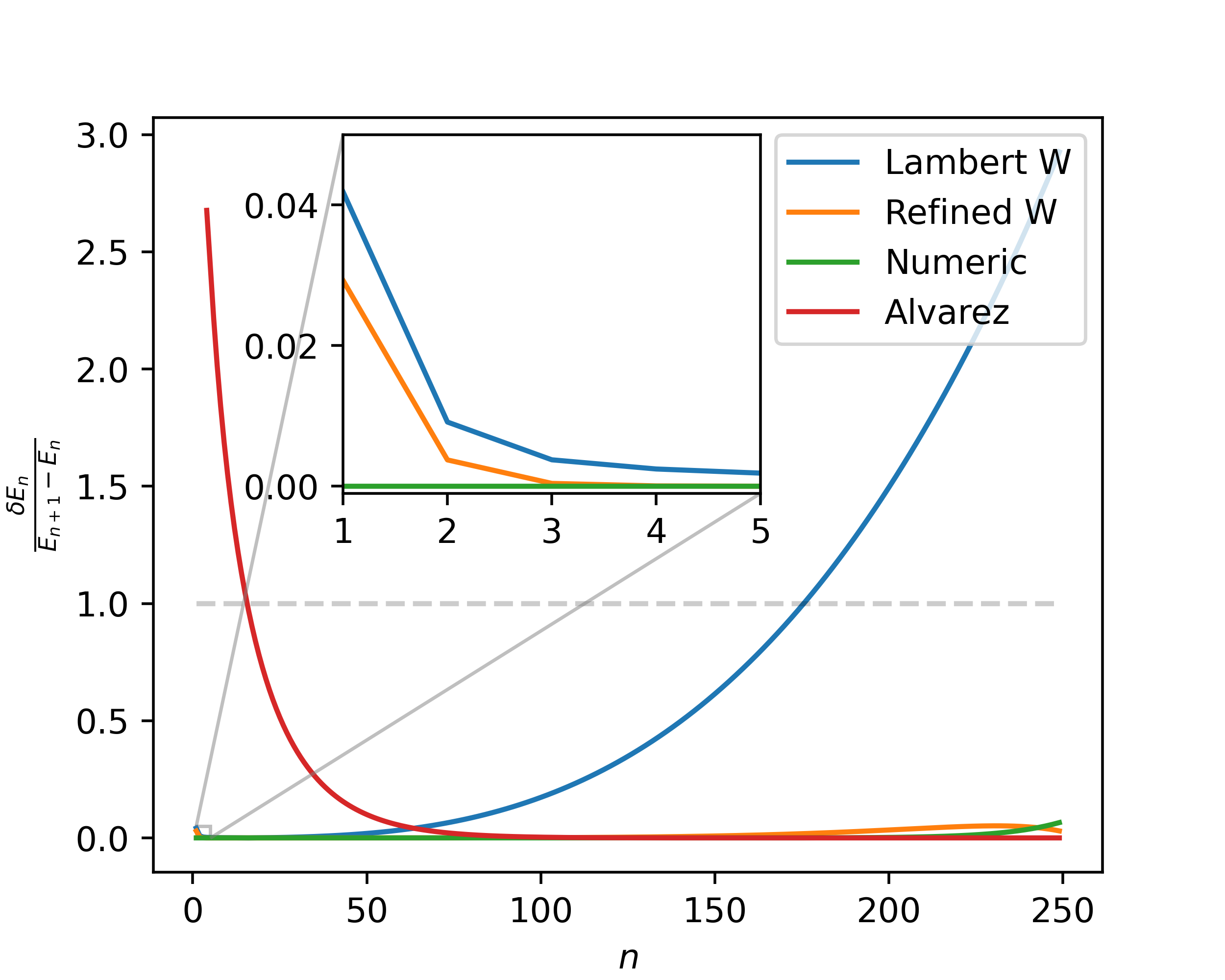}
    \caption{
Absolute errors of the approximate positive energy levels normalized by
the spacing between adjacent exact levels at $N=1000$. The blue curve
shows the leading Lambert-$W$ approximation~\eqref{eq:normal_small_E},
the orange curve shows its refined version~\eqref{eq:intermediate_t_action_correction},
the green curve corresponds to the numerical solution of the implicit
quantization condition~\eqref{eq:bleher_without_zero_part}, and the red
curve shows the Alvarez approximation~\eqref{eq:alvarez_full}. The
refined and numerical approximations use the extended expansion of
$\mathcal R(h,t)$ from \eqref{eq:I_all_terms}. The small-$t$
approximation~\eqref{eq:ultrasmall_E} is omitted because, for $N=1000$,
it is accurate only for the first positive level and rapidly loses
accuracy as $n$ increases.
}
    \label{pic:relative_accuracy}
\end{figure}

Figure \ref{pic:evals} shows the exact energy levels obtained by matrix diagonalization. To compare the approximations quantitatively, we consider the absolute error normalized by the spacing between adjacent exact levels. This quantity is shown in Fig. \ref{pic:relative_accuracy} for the implicit separatrix quantization condition, the explicit large-$t$ approximation, its refined version, and the Alvarez approximation.

\section{Conclusion}\label{sec:conclusion}

We have studied the energy spectrum of the degenerate $\chi^{(2)}$ Hamiltonian in the limit of a large total excitation number. By mapping the spectral problem onto an effective one-dimensional Schrödinger equation, we applied Bleher's separatrix quantization condition to the levels located near the center of the optical spectrum.

After taking into account the parity restriction and the indexing of the physical energy levels, we obtained the implicit quantization condition \eqref{eq:bleher_without_zero_part}. Its expansion in the regimes $t\to0$ and $t\to\infty$ with $ht\to0$ yields the two explicit asymptotic expressions \eqref{eq:ultrasmall_E} and \eqref{eq:normal_small_E}, respectively. The first describes the energy levels closest to zero, whereas the second applies to a broader intermediate range. In the transition region $t=O(1)$, the implicit condition can be solved numerically.

Comparison with exact matrix diagonalization shows that the separatrix quantization condition accurately describes the near-zero-energy part of the spectrum, where the conventional Alvarez approximation is less accurate. Together, the present formulas and the Alvarez approximation provide complementary descriptions of the spectrum from the central levels to the vicinity of its edges.

The obtained energy-level asymptotics may serve as a basis for approximating the corresponding eigenvectors and, subsequently, the quantum dynamics of second-harmonic generation and parametric down-conversion at large excitation numbers.
\section*{Acknowledgments}

The author is grateful to Valentin Gorshenin for valuable discussions and to Evgeny Vyborny for bringing Bleher's work to his attention. The research was carried out with the financial support of the Russian Science Foundation within the framework of project No. 25-12-00263.

\appendix
\section{Asymptotics of $I(\varepsilon)$}\label{appendix_I_asymptotics}
The main idea of finding the asymptotics is to find $I(\varepsilon)$ as an integral of its derivative:
\begin{eqw}\label{eq:I_expansion_idea}
    I(\varepsilon) = I(0) + \int\limits_{0}^{\varepsilon} \frac{\partial I(\varepsilon')}{\partial \varepsilon'}d\varepsilon'
\end{eqw}
First, let's replace $u=x^2$ in the expression \eqref{eq:I_def}:
\begin{eqw}
    I(\varepsilon) = \int\limits_{u_{\min}}^{u_{\max}}\sqrt{-\varepsilon - 4u^3+3u} \frac{du}{2\sqrt{u}}
\end{eqw}
The integration limits are determined by the two positive roots of the cubic equation:
\begin{eqw}
    3u-4u^3 = \varepsilon
\end{eqw}
Using the triple angle sine formula, the solution can be represented as:
\begin{eqw}\label{eq:u_l_definition}
    u_l = \sin\br{\frac{1}{3}\br{2\pi l + \arcsin\varepsilon}}
\end{eqw}
The case $l=-1$ corresponds to the negative root $u_{-}$, the other two cases correspond to the integration limits: $u_0 = u_{\min}$ and $u_{+} =u_{\max}$. In the limit $\varepsilon = 0$, $u_0 = 0$ and $u_+ = \sqrt{3}/2$. Now it is easy to see that for $\varepsilon=0$ the integral is taken exactly:
\begin{eqw}
    I(0) = \int\limits_{0}^{\sqrt{3}/2}\sqrt{3/4-u^2} du = \frac{3\pi}{16}
\end{eqw}

The derivative of $I(\varepsilon)$ is given by
\begin{eqw}
    \frac{\partial I}{\partial \varepsilon} = -\int\limits_{u_0}^{u_+}\frac{du}{4\sqrt{u(-\varepsilon - 4u^3+3u)}}
\end{eqw}
Next, we factor the denominator into linear factors and find the expression of the integral in terms of a complete elliptic integral of the first kind \cite{gradshteyn2014table}:
\begin{eqw}\label{eq:I_trough_K}
    \frac{\partial I}{\partial \varepsilon} = -\int\limits_{u_0}^{u_+}\frac{du}{8\sqrt{u(u-u_-)(u-u_0)(u_+-u)}} = -\frac{1}{4\sqrt{u_+(u_0 - u_-)}}K\br{\sqrt{\frac{\br{u_+ - u_0}\br{-u_-}}{u_+ \br{u_0 - u_-}}}},
\end{eqw}

Next, we can apply the formula from \cite{olver2010nist}:
\begin{eqw}\label{eq:NIST_expansion}
    K\br{\sqrt{1-k'^2}} &= \sum\limits_{m=0}^{\infty} \br{\frac{(1/2)_m}{m!}}^2k'^{2m}\br{\ln(1/k')+2\br{H_{m} - H_{2m} + \ln(2)}},
\end{eqw}
where $H_m$ are harmonic numbers.
Substituting \eqref{eq:NIST_expansion} into \eqref{eq:I_trough_K}, and then using \eqref{eq:u_l_definition}, we obtain the small-$\varepsilon$ expansion of $I'(\varepsilon)$. Substitution of this expansion into \eqref{eq:I_expansion_idea} yields:
\begin{eqw}\label{eq:I_all_terms}
    I(\varepsilon) &= \frac{3 \pi }{16} + 3\br{\frac{\varepsilon}{12\sqrt{3}}}^1
\left[\ln\br{\frac{\varepsilon}{12\sqrt{3}}} - 1\right]+ \\
&+
4\left(\frac{\varepsilon}{12\sqrt3}\right)^3
\left[
34+15\ln\left(\frac{\varepsilon}{12\sqrt3}\right)
\right]
\\
&+
\frac{54}{5}
\left(\frac{\varepsilon}{12\sqrt3}\right)^5
\left[
2003+770\ln\left(\frac{\varepsilon}{12\sqrt3}\right)
\right]
\\
&+
\frac{24}{7}
\left(\frac{\varepsilon}{12\sqrt3}\right)^7
\left[
1394891+510510\ln\left(\frac{\varepsilon}{12\sqrt3}\right)
\right]
\\
&+
\frac{1}{3}
\left(\frac{\varepsilon}{12\sqrt3}\right)^9
\left[
3751204337
+1338557220\ln\left(\frac{\varepsilon}{12\sqrt3}\right)
\right]
\\
&+
\frac{108}{55}
\left(\frac{\varepsilon}{12\sqrt3}\right)^{11}
\left[
184132249871
+64696932300\ln\left(\frac{\varepsilon}{12\sqrt3}\right)
\right] + O\br{\varepsilon^{13} \ln\varepsilon}.
\end{eqw}
\section{Simplification of equation \eqref{eq:bleher_eq}}\label{appendix_gamma_simplification}
Let’s start by simplifying the left‑hand side of equation \eqref{eq:bleher_eq}. Using the notation \eqref{eq:h_def} for $h$:
\begin{eqw}
    \frac{I(2\sqrt{3}ht)}{h} = \pi\left(\frac N4+\frac38\right)+
\frac{t}{2}
\left(\ln h+\ln t-1-\ln 6 \right)
+\mathcal R(h,t)
\end{eqw}
Now let’s move on to the right‑hand side of the equation. We apply Legendre’s duplication formula:
\begin{eqw}
    \Gamma(z)\Gamma\left(z+\frac12\right)
=
2^{1-2z}\sqrt{\pi}\,\Gamma(2z).
\end{eqw}

At
$
z=\frac14-\frac{it}{2}
$
we get

\begin{eqw}
    \Gamma\left(\frac14-\frac{it}{2}\right)
\Gamma\left(\frac34-\frac{it}{2}\right)
=
2^{\frac12+it}\sqrt{\pi}\,
\Gamma\left(\frac12-it\right),
\end{eqw}

where we get

\begin{eqw}
    \arg\Gamma\left(\frac14-\frac{it}{2}\right)
+
\arg\Gamma\left(\frac34-\frac{it}{2}\right)
=
t\ln2+
\arg\Gamma\left(\frac12-it\right).
\end{eqw}

Further from the reflection formula

\begin{eqw}
    \Gamma(z)\Gamma(1-z)=\frac{\pi}{\sin\pi z}
\end{eqw}

for the same \(z\), we have

\begin{eqw}
    \arg\Gamma\left(\frac14-\frac{it}{2}\right)
-
\arg\Gamma\left(\frac34-\frac{it}{2}\right)
=
-\arg\sin\left(\frac{\pi}{4}-\frac{i\pi t}{2}\right).
\end{eqw}

Since
\begin{eqw}
    -\arg\sin\left(\frac{\pi}{4}-\frac{i\pi t}{2}\right)
=
\arctan e^{\pi t}-\frac{\pi}{4}.
\end{eqw}

By adding or subtracting the two obtained relations, depending on the parity of \(N\), we get
\begin{eqw}
    2\arg\Gamma
\left(
\frac12-\frac{(-1)^N}{4}-\frac{it}{2}
\right)
=
t\ln2+
\arg\Gamma\left(\frac12-it\right)
+
(-1)^N
\left(
\arctan e^{\pi t}-\frac{\pi}{4}
\right).
\end{eqw}
Finally, using the obtained relations for the gamma function and the asymptotic $I(\varepsilon)$:
\begin{eqw}
    \pi
\left(
n+\frac N4+\frac38-\widetilde n
-\frac38(-1)^N
\right)+
\mathcal R(h,t)= \frac t2\ln\frac3h
+
\arg\Gamma
\left(
\frac12-\frac{(-1)^N}{4}-\frac{it}{2}
\right).
\end{eqw}
\section{Low-energy expansion}\label{appendix_small_energies}
The error estimates in both appendices concern the inversion of
Eq.~\eqref{eq:bleher_without_zero_part}; the accuracy of the underlying
semiclassical quantization is a separate contribution. Appendix~\ref{appendix_accuracy} combines these errors and compares them with the adjacent-level spacing.

For this derivation, write $a_N=\frac12-\frac{(-1)^N}{4}$, so that
$D_N=\ln(3/h)-\psi(a_N)$. Taylor expansion at $t=0$ gives
\begin{eqw}\label{eq:small_gamma_series}
    \arg\Gamma\left(a_N-\frac{it}{2}\right)
    =-\frac{\psi(a_N)}{2}t+\frac{\psi^{(2)}(a_N)}{48}t^3+O(t^5).
\end{eqw}
The Taylor series converges for $|t|<2a_N$, the distance to the nearest
pole. Substitution into Eq.~\eqref{eq:bleher_without_zero_part} yields
\begin{eqw}\label{eq:small_phase_with_remainder}
    \frac{D_N}{2}t+\frac{\psi^{(2)}(a_N)}{48}t^3
    =\pi\kappa_n+\mathcal R(h,t)+O(t^5).
\end{eqw}
For $h\to0$ and $\frac12\leq\kappa_n=o(D_N)$, put
$\tilde t_n=2\pi\kappa_n/D_N$. Then $\tilde t_n\to0$ and
$|\ln(h\tilde t_n)|=O(D_N)$. Equation~\eqref{eq:small_phase_with_remainder}
first gives $t=\tilde t_n+O(\tilde t_n^3/D_N)$; substituting this estimate
back into its nonlinear terms gives
\begin{eqw}\label{eq:small_t_action_correction}
    t=\tilde t_n-\frac{\psi^{(2)}(a_N)\tilde t_n^3}{24D_N}
    +\frac{2\mathcal R(h,\tilde t_n)}{D_N}
    +O\left(\frac{\tilde t_n^5}{D_N}\right).
\end{eqw}
By Eq.~\eqref{eq:I_asymptotics_till_5}, the action contribution has
order $O(h^2\kappa_n^3/D_N^3)$, which is
$o(\kappa_n^5/D_N^6)$ since $h^2D_N^3/\kappa_n^2\to0$.
Consequently,
\begin{eqw}\label{eq:small_t_explicit}
    t=\frac{2\pi\kappa_n}{D_N}
    -\frac{\psi^{(2)}(a_N)(\pi\kappa_n)^3}{3D_N^4}
    +O\left(\frac{\kappa_n^5}{D_N^6}\right).
\end{eqw}
Multiplying by $\sqrt{2N+3}$ gives Eq.~\eqref{eq:ultrasmall_E}.
Since $D_N\sim\ln N$, the assumed range is $\kappa_n=o(\ln N)$.

\section{Intermediate-energy expansion}\label{appendix_medium_energies}
We consider the joint limit
\begin{eqw}\label{eq:intermediate_regime}
    h\to0,\qquad t\to\infty,\qquad ht\to0.
\end{eqw}
With $a_N$ as in Appendix~\ref{appendix_small_energies}, the shifted
Stirling expansion \cite{olver2010nist} gives
% Reference: https://dlmf.nist.gov/5.11.E8
\begin{eqw}\label{eq:intermediate_gamma_series}
    \arg\Gamma\left(a_N-\frac{it}{2}\right)
    =\frac t2\left(1-\ln\frac t2\right)+\frac\pi8(-1)^N
    -\frac{1}{48t}+O(t^{-3}).
\end{eqw}
Here the coefficient of $t^{-1}$ is
$a_N^2-a_N+1/6=-1/48$ for both parities of $N$.
For this appendix, abbreviate
\begin{eqw}\label{eq:zeta_def}
    \zeta_n=\pi\kappa_n-\frac\pi8(-1)^N.
\end{eqw}
Equation~\eqref{eq:bleher_without_zero_part} becomes
\begin{eqw}\label{eq:intermediate_phase_with_remainder}
    \frac t2\ln\frac{6e}{ht}
    =\zeta_n+\mathcal R(h,t)+\frac{1}{48t}+O(t^{-3}).
\end{eqw}
Omitting the remainder terms and setting $w=-\ln(6e/(ht))$ gives
$we^w=-h\zeta_n/(3e)$ and $t=-2\zeta_n/w$.
Since $ht\to0$ requires $w\to-\infty$, we choose the branch $W_{-1}$:
\begin{eqw}\label{eq:intermediate_t_leading}
    t\approx\tau_n=\frac{2\zeta_n}{-W_{-1}\left(-\dfrac{h\zeta_n}{3e}\right)}.
\end{eqw}
Multiplication by $\sqrt{2N+3}$ gives Eq.~\eqref{eq:normal_small_E}.

Expanding Eq.~\eqref{eq:intermediate_phase_with_remainder} about $t=\tau_n$ and retaining the leading corrections, we obtain:
\begin{eqw}\label{eq:intermediate_t_action_correction}
    t=\tau_n+\frac{2\mathcal R(h,\tau_n)}{\ln(6/(h\tau_n))}
    +\frac{1}{24\tau_n\ln(6/(h\tau_n))}
    +O\left(h^4\tau_n^5+\frac{1}{\tau_n^3\ln(6/(h\tau_n))}\right).
\end{eqw}
The error includes the variation of the retained corrections and the
next Stirling term; the mixed term $O(h^2\tau_n/\ln(6/(h\tau_n)))$ is absorbed
by the displayed bound.

As $h\zeta_n\to0$, together with $h\sim3/(4N)$ and $\zeta_n\sim\pi n$, this gives
$\tau_n\sim2\pi n/\ln(N/n)$. Thus the joint limit
\eqref{eq:intermediate_regime} corresponds to $\ln N\ll n\ll N$
and yields Eq.~\eqref{eq:intermediate_scaling}.

\section{Accuracy of the explicit approximations}\label{appendix_accuracy}
We first estimate the error in approximating a solution of
Eq.~\eqref{eq:bleher_without_zero_part}, and then include the error of
semiclassical quantization. Let $\delta t$ denote the approximation error
and $\Delta t$ the spacing between adjacent positive levels. Since
$E_n^N=\sqrt{2N+3}\,t$, the ratio $|\delta t|/\Delta t$ is also the energy
error normalized by the adjacent-level spacing.

For $\frac12\leq\kappa_n=o(D_N)$, the derivative of the quantization
phase is asymptotic to $D_N/2$. Thus Eq.~\eqref{eq:small_t_explicit} gives
\begin{eqw}\label{eq:small_accuracy}
    \delta t=O\left(\frac{\kappa_n^5}{D_N^6}\right),\qquad
    \Delta t\sim\frac{2\pi}{D_N},\qquad
    \frac{|\delta t|}{\Delta t}=O\left(\frac{\kappa_n^5}{D_N^5}\right)=o(1).
\end{eqw}

In the intermediate regime, differentiation of the leading phase in
Eq.~\eqref{eq:intermediate_phase_with_remainder} gives
$\frac12\ln(6/(ht))$. The elliptic expression for $I'$ also yields
$\partial_t\mathcal R(h,t)=O(h^2t^2|\ln(ht)|)$, which changes this
phase derivative by a relative $O((ht)^2)$. Consequently,
\begin{eqw}\label{eq:intermediate_spacing}
    \Delta t\sim\frac{2\pi}{\ln(6/(h\tau_n))}.
\end{eqw}
For the leading approximation \eqref{eq:intermediate_t_leading},
Eq.~\eqref{eq:intermediate_t_action_correction} implies
\begin{eqw}\label{eq:intermediate_leading_accuracy}
    \delta t&=O\left(h^2\tau_n^3+\frac{1}{\tau_n\ln(6/(h\tau_n))}\right),\\
    \frac{|\delta t|}{\Delta t}
    &=O\left(h^2\tau_n^3\ln\frac6{h\tau_n}+\frac1{\tau_n}\right).
\end{eqw}
The relative error $|\delta t|/\tau_n$ tends to zero throughout
$\tau_n\to\infty$, $h\tau_n\to0$. An error smaller than the adjacent-level
spacing requires the stronger condition
\begin{eqw}\label{eq:intermediate_spacing_condition}
    h^2\tau_n^3\ln\frac6{h\tau_n}\to0.
\end{eqw}

Retaining the two explicit corrections in
Eq.~\eqref{eq:intermediate_t_action_correction} gives instead
\begin{eqw}\label{eq:intermediate_corrected_accuracy}
    \delta t&=O\left(h^4\tau_n^5+\frac{1}{\tau_n^3\ln(6/(h\tau_n))}\right),\\
    \frac{|\delta t|}{\Delta t}
    &=O\left(h^4\tau_n^5\ln\frac6{h\tau_n}+\frac1{\tau_n^3}\right).
\end{eqw}
The same bound holds if $\mathcal R(h,\tau_n)$ is replaced by its explicit
cubic term in Eq.~\eqref{eq:I_asymptotics_till_5}. The corresponding
sufficient condition for an error smaller than the spacing is
\begin{eqw}\label{eq:intermediate_corrected_spacing_condition}
    h^4\tau_n^5\ln\frac6{h\tau_n}\to0.
\end{eqw}
This is weaker than Eq.~\eqref{eq:intermediate_spacing_condition}, since
$h\tau_n\to0$.

Finally, within the range of Bleher's $O(h^2)$ estimate for the scaled
energy $\varepsilon_n$ \cite{bleher1994semiclassical}, the underlying
quantization contributes
\begin{eqw}\label{eq:quantization_error_scaling}
    \delta\varepsilon=O(h^2),\qquad \delta t=O(h),\qquad
    \delta E=O(h^{1/2}).
\end{eqw}
Its contribution to the error normalized by the spacing is $O(hD_N)$
in Eq.~\eqref{eq:small_accuracy} and
$O(h\ln(6/(h\tau_n)))$ in
Eqs.~\eqref{eq:intermediate_leading_accuracy} and
\eqref{eq:intermediate_corrected_accuracy}. Both tend to zero in their
respective limits. The total error is bounded by the sum of the
inversion and quantization contributions.

\bibliography{references}
\end{document}